\documentstyle[preprint,aps]{revtex}
\tightenlines

\begin{document}

\draft
\preprint{INJE-TP-01-06, hep-th/0107065}

\title{ Quintessence and Brane world scenarios}

\author{ Y.S. Myung\footnote{Email-address :
ysmyung@physics.inje.ac.kr}}
\address{Department of Physics, Graduate School, Inje University, Kimhae 621-749, Korea}

\maketitle

\begin{abstract}
We discuss  the possibility of quintessence in the dilatonic
domain walls including the Randall-Sundrum brane world.
We obtain the zero mode effective action for
gravitating objects in the dilatonic domain wall.
First we consider the four dimensional (4D) gravity and the Brans-Dicke  graviscalar  with
a potential. This  can be further rewritten as a minimally
coupled scalar with the Liouville-type potential in the Einstein frame.
However this model fails to induce the quintessence on the
dilatonic domain wall because the potential is negative.
 Second we consider the 4D gravity with the
dilaton. In this case we find also a negative potential.
 Any negative potential  gives  us
negative energy density and positive pressure, which does not
lead to an accelerating universe.
Consequently it turns out that the zero mode approach of the dilatonic
domain wall cannot accommodate the quintessence in cosmology.

\end{abstract}
\vfill
Compiled at \today : \number \time.

\newpage
\section{Introduction}
\label{introduction}
Recently cosmic quintessence has proposed to be an alternative way
to understand the astronomical data of supernova which indicate
that the universe is presently accelerating\cite{Per,CDS,Gar}. Also this may be considered
as a method to resolve the cosmological constant problem.
This is possible because instead of the fine-tuning, it provides a
 model of slowly decaying cosmological constant.
Several literature  discussed on this issue within the string
theory\cite{Wit,HKS} and  cosmological context\cite{RP,KLa}.

More recently authors in ref.\cite{CL} also considered this problem in
the  dilatonic domain wall. After the integration over the dual holographic
field theory, they obtained a dilatonic gravity with the potential on the brane.
They used
the Hamilton-Jacobi method inspired by the holographic
renormalization group to investigate the intrinsic
Friedmann-Robertson-Walker (FRW) cosmology on the brane.
It was shown that the holographic
quintessence is allowed on the dilatonic brane
because a Liouville-type potential appears on the brane.
This case  requires a negative slowly varying bulk potential
which implies that
the bulk space is a five dimensional anti de Sitter (AdS$_5)$ spacetime.

In this paper we wish to deal with the same issue within a different
context.
An idea of  the brane world scenario is that our universe may be a brane embedded in
the higher dimensional space\cite{RS,Aka,RS1,RS2,ADDK}. A concrete model is a single 3-brane embedded
in the AdS$_5$ space\cite{RS2}.  Randall and
Sundrum (RS) have shown that a longitudinal part ($h_{\mu\nu}$) of
the
metric fluctuations satisfies the Schr\"odinger-like equation with an
attractive delta-function. As a result, the massless Kaluza-Klein (KK) modes which
describe the localized gravity on the brane were found. Furthermore,
the massive KK modes which reside in the extra dimension
lead to  corrections to the Newtonian
potential.
We would like to point out that this has been done
in the 4D Minkowski
brane with the RS gauge\footnote{In fact, this gauge for the 5D metric fluctuation $h_{MN}$ is
composed of Gaussian-normal  gauge ($h_{44} = h_{4\mu}=0$)
and 4D transverse, traceless  gauge ($\partial^\mu h_{\mu\nu} =0$,
$h^\mu_{~\mu} = 0$).}.
It seems that this gauge is so restrictive.
In order to
have  an interesting cosmological model on the brane, we may include
the graviscalar (KK scalar) at the beginning\cite{Iva,MyuK,MyuKL}.

In this work we will not use the
holographic renormalization group  to find  the  brane potential.
Instead  we follow an idea for the genuine brane world scenario such
that the  universe is filled with the zero
 modes of the bulk fields which are trapped on the
 brane by the gravitational interaction\cite{BG}. In our approach the relevant scalar
 is either the dilaton ($d$) or the graviscalar $(h)$.
 The condition that the
 zero mode is localized on the brane corresponds  to the
normalizability of the ground state wave function on the brane.
Hence we  have to find the finite zero-mode  action of  the dilatonic domain wall to
study its  cosmological implication. Fortunately we can obtain the
finite action with  the Liouville-type potentials.  But we fail to find any
accelerating universe from the dilatonic domain wall.

The organization of our paper is as follows. In Sec. II we briefly
review a simple model for the quintessence. We derive the 4D
effective action on the brane by using the zero-mode approach
in Sec. III. In Sec. IV we consider the graviscalar as a dynamic scalar on the
brane. And we discuss its role for cosmological implication.
In Sec. V we consider the dilaton as a dynamic scalar for deriving a cosmological
evolution. Finally we discuss our results in Sec. VI.

\section{Quintessence}
Quintessence endeavors to handle a cosmological problem with a
dynamic negative pressure. A minimally coupled
scalar field with a potential that decreases as the field
increases is usually introduced for this purpose\cite{RP}.
 The  action is given by
\begin{equation}
S_{\rm Q} = {1 \over {2 \kappa_4^2}} \int d^4 x \sqrt{-g} \left [
 R - (\partial \phi)^2 -2 V(\phi)\right]
\label{Qact}
\end{equation}
with the 4D gravitational constant $\kappa_4^2$.
This is a canonically normalized scalar action coupled to the 4D gravity.
The Einstein equation is
\begin{equation}
R_{\mu\nu}- \frac{1}{2}R g_{\mu\nu}= T_{\mu\nu}
\label{Qeq}
\end{equation}
with
\begin{equation}
T_{\mu\nu}= \partial_\mu\phi \partial_\nu\phi
-\frac{1}{2}(\partial \phi)^2g_{\mu\nu} -V(\phi)g_{\mu\nu}.
\label{Qst}
\end{equation}
Considering the FRW flat metric of $ds^2_{FRW}=-dt^2
+{\cal R}^2(t)dx^idx_i$,  the equation of motion for $\phi$ and the
conservation law of $\nabla_\mu T^{\mu0}=0$ lead to the the same
equation as $\ddot \phi + 3\frac{\dot {\cal R}}{{\cal R}} \dot \phi +
V'(\phi)=0$. The two FRW equations are given by
\begin{equation}
\frac{\dot {\cal R}^2}{{\cal R}^2} = \frac{\rho}{3},~~~ \frac{\ddot {\cal R}}{{\cal R}}
=-\frac{\rho+3p}{6}.
\label{Eins}
\end{equation}
Assuming $\phi=\phi(t)$ for cosmological purpose, then the energy
density and pressure are given by
\begin{equation}
\rho= \frac{1}{2} \dot \phi^2 +V(\phi),~~~~p= \frac{1}{2} \dot \phi^2
-V(\phi).
\label{Qep}
\end{equation}
The corresponding equation of state takes the form
\begin{equation}
\omega \equiv \frac{p}{\rho}= \frac{ \dot \phi^2 -2V(\phi)}{ \dot \phi^2 +2V(\phi)}.
\label{Qeqs}
\end{equation}
The equation of state  ranges over $-1<\omega<1$, depending on the
dynamics of the field.
We note that when $ \dot \phi^2 <V(\phi)$ on later time,
  an accelerating universe appears from the second equation in Eq.(\ref{Eins}).
In this model  $\phi$ and $\rho$ scale
as\cite{RP}
\begin{equation}
\frac{\partial \phi}{\partial {\cal R}}= \frac{\sqrt {3(1+\omega)}}{{\cal R}},~~~
\rho \sim \frac{1}{{\cal R}^{3(1+\omega)}}.
\end{equation}
Using the relation of $V(\phi)=(1-\omega) \rho/2$ together with
 $\phi=\sqrt{3(1+\omega)}\ln {\cal R}$,
 the positive potential takes the form
\begin{equation}
V_{quint}(\phi)= V_0 e^{-\sqrt{3(1+\omega)} \phi},~~~3(1+\omega)<2
\label{Qpot}
\end{equation}
which is a kind of Liouville-type potential that decreases as $\phi$ increases.
According to the theory of quintessence, the dark energy of the
universe is dominated by the scalar potential
which is still rolling to its minimum of $V_{quint}=0$. We
require its minimum at $\phi=\infty$ conventionally. The above
potential is suited well  for the quintessence. For example, if
one takes $\omega=-1/2<-1/3$, $V_{quint}(\phi)=V_0 e^{-\sqrt{3/2} \phi}$
can induce an accelerating universe.
In the next section we wish to check whether or not the above-type potential can be
found from the dilatonic domain wall model.

\section{ 4D effective action on the brane}
\label{sec-randall}

We start with the 5D bulk action and
4D domain wall action  as\cite{You0001166,DDW}
\begin{equation}
 S = S_{\rm bulk} + S_{\rm DW}
\label{action}
\end{equation}
with
\begin{eqnarray}
S_{\rm bulk} &=& {1 \over { 2 \kappa_5^2}} \int d^5x \sqrt{-G}
  \left [ R_5 - {4 \over 3} \partial_M D \partial^M D
  - e^{-2 a D} \Lambda \right ],
\label{actionbulk} \\
S_{DW} &=& - \sigma_{\rm DW} \int d^4 x \sqrt{-\gamma} e^{-a D} ,
\label{actiondw}
\end{eqnarray}
where $\sigma_{\rm DW}$ is the tension of the domain wall and $\gamma$
is the determinant of the  induced metric
$\gamma_{\mu\nu} = \partial_\mu X^M \partial_\nu X^N G_{MN}$ on  the
domain wall.
Here $M,N=0,1,2,3,4(x^4=z)$ and $\mu, \nu = 0,1,2,3(x^\mu = x)$.
``$D$'' denotes the dilaton.
We are interested in the dilatonic domain wall solution with\footnote{For $a=0$ case,
this reduces to  the second RS
vacuum solution exactly\cite{ML}.}

\begin{equation}
{\bar G}_{MN} = H^{-2}(z) \eta_{MN}, ~~ e^{2 \bar D}=[H(z)]^{\frac{9a}{4}},
~~ \Lambda = \frac{32 k^2}{\Delta}, ~~ \sigma_{\rm DW} = \frac{16k
}{|\Delta|\kappa_5^2}
\label{rssolution}
\end{equation}
with the conformal factor $H(z)= (1+ 4k \frac{\Delta +2}{\Delta}|z|)^{4/(3\Delta +6)},
 \Delta=-\frac{8}{3}+\frac{3a^2}{2}$
 and $\eta_{MN} = {\rm diag}[-++++]$.
Here overbar($^-$) means the vacuum solution. Also we choose a
negative bulk cosmological constant $\Lambda<0$ with $\Delta<0$ and a positive
domain wall tension $\sigma_{\rm DW}>0$ for the fine-tuning. The condition of $\Delta<0$
leads to the constraint on $a$ : $0<a^2<16/9$.

In order to obtain  the zero mode effective action, we propose the
metric  $G_{MN} = H^{-2}(z) {\hat G}_{MN}$ where ${\hat G}_{MN}$
is a function of ``$x$" only.
Explicitly the line element is given by
\begin{eqnarray}
dS_5^2 &=& G_{MN} dx^M dx^N = H^{-2} {\hat G}_{MN} dx^M dx^N
\nonumber \\
   &=& H^{-2} \left [ g_{\mu\nu}(x) dx^\mu dx^\nu
    + h(x)^2 dz^2 \right ].
\label{5metric}
\end{eqnarray}
Further we introduce the dilaton $d(x)$ in the form of
\begin{equation}
e^{2  D}=[H(z)]^{\frac{9a}{4}}e^{2d(x)}.
\label{dilaton}
\end{equation}
Hereafter the graviton $g_{\mu\nu}(x)$, graviscalar $h(x)$, and dilaton $d(x)$
play the role of zero modes.
Off-diagonal elements are not included because they are not necessary here
\footnote{One may introduce  off-diagonal
term
of $2A_\mu(x)dx^\mu dz$ for general discussion\cite{KM}.}.
$h(x)$ is related to the radion that
is necessary for stabilizing the distance between two branes
 in the first RS model\cite{RS1}.
Substituting Eq.(\ref{5metric}) together with Eq.(\ref{dilaton}) into (\ref{action}) and
then integrating it over $z$ lead to the 4D effective action
\begin{equation}
S_{\rm DDW1} = {1 \over {2 \kappa_5^2}} \int d^4 x \sqrt{-g} \left [
-\frac{\Delta}{2k(\Delta+4)} \left(h R -\frac{4}{3} h (\partial d)^2 \right)
- \frac{16k}{\Delta}\left ( h e^{-2ad} + { 1\over h} -2 e^{-ad} \right ) \right
].
\label{act1}
\end{equation}
If we define the 4D gravitational constant as
\begin{equation}
\kappa^2_4=-2k \frac{\Delta+4}{\Delta} \kappa^2_5,
\end{equation}
then the above action leads to
\begin{equation}
S_{\rm DDW2} = {1 \over {2 \kappa_4^2}} \int d^4 x \sqrt{-g} \left [
 h R -\frac{4}{3} h (\partial d)^2 -  \frac{32k^2(\Delta +4)}{\Delta^2}\left (2e^{-ad}
 - h e^{-2ad} - { 1\over h}\right ) \right
].
\label{act2}
\end{equation}
Here $\kappa^2_4>0$ is automatically guaranteed because  we choose $\Delta<0$ and $ \Delta+4>0$.
If $\Delta>0$, the 4D gravity is not trapped on the dilatonic
domain wall\cite{DYoum}. This action is very important for our study.
 We note here that the limit of $k \to 0$ means that
AdS$_5$ space $\to$ 5D Minkowski space, and  tension domain wall $\to$
tensionless domain wall. In this limit we recover the conventional
KK model without the last potential in Eq.(\ref{act2}).

\section{Brans-Dicke graviscalar as a dynamic scalar}

Since the above action has two different scalars with the mixed potential, we first
consider the dynamic graviscalar  $(d(x)=0,h(x)\not=1)$. In the next section
we will deal with the dilaton. The action
(\ref{act2}) reduces to the Brans-Dicke (BD) model with a
different
potential\cite{You0001166}

\begin{equation}
S_{\rm BD} = {1 \over {2 \kappa_4^2}} \int d^4 x \sqrt{-g} \left [
 h R -  \frac{32k^2(\Delta +4)}{\Delta^2}\left (2- h - { 1\over h}\right ) \right
].
\label{BD}
\end{equation}
The first term ($h R$) is
 the BD term when the BD parameter $w=0$\cite{Bra61PR925}.
This comprises  the massless Kaluza-Klein
modes  $g_{\mu\nu}$, $g_{44}(\sim h^2)$ but with
$g_{\mu4}(\sim A_\mu)=0$~\footnote{
This model is equivalent to
$S_{\rm KK} = {1 \over 2 \kappa_5^2} \int d^5 x \sqrt{-G} R_5$ with
 a factorizable geometry of $H=1(k=0)$.}.
In this sense, we wish to call $h$ the BD scalar.
The second term  arises from the facts : a dilatonic domain wall is located at
$z=0$$(``2")$,
the bulk spacetime is an  AdS$_5$ space $(``h")$, and $h$ is  the KK scalar$(``1/h")$.
Equivalently, this means that the domain wall configuration describes
non-factorizable geometry with a conformal factor $H(z)$.
Especially for $a=0$ case, we have the effective action for the RS-type model\cite{ML}
: ${\cal L}_{\rm RS} =\sqrt{-g} [h R - 6 k^2(2-h + 1/h )]/2\kappa^2_4$.

In order to obtain a canonical scalar action, we use a conformal
transform :  $g_{\mu\nu} \to \Omega^{-2} \bar g_{\mu\nu}$ with
$\Omega^2=h$. This implies that we move from the string-like frame to the
Einstein frame. The resulting action takes the form
\begin{equation}
S_{\rm BD1} = {1 \over {2 \kappa_4^2}} \int d^4 x \sqrt{-\bar g} \left [
  R -\frac{3}{2}\frac{(\partial h)^2}{h^2}-  \frac{32k^2(\Delta +4)}
  {\Delta^2}\frac{1}{h^2}\left (2- h - 1/h\right ) \right].
\label{act3}
\end{equation}
Let us define $h\equiv e^{\sqrt{2/3} \Phi}$ to obtain a canonical form like (\ref{Qact}).
Then we have a desired action
\begin{equation}
S_{\rm BD2} = {1 \over {2 \kappa_4^2}} \int d^4 x \sqrt{-\bar g} \left [
  R -(\partial \Phi)^2- \frac{32k^2(\Delta +4)}
  {\Delta^2}\left (2e^{-\sqrt{8/3} \Phi}-e^{-\sqrt{6} \Phi}-
  e^{-\sqrt{2/3} \Phi}\right ) \right]
\label{act4}
\end{equation}
which implies the graviscalar potential on the  domain
wall
\begin{equation}
V_{GS}(\Phi)=  \frac{16k^2(\Delta +4)}
  {\Delta^2}\left (2e^{-\sqrt{8/3} \Phi}-e^{-\sqrt{6} \Phi}-
  e^{-\sqrt{2/3} \Phi}\right).
  \label{dddp}
\end{equation}
All of terms belong to the Liouville-type
potential.
The only last term that comes from the genuine
graviscalar satisfies a criterion of the quintessence
($\sqrt{2/3}<\sqrt{2}$).
Unfortunately it belongs to a negative potential.
For this  purpose we rewrite the potential as

\begin{equation}
V_{GS}(\Phi)= - \frac{16k^2(\Delta
  +4)}{\Delta^2}e^{-\sqrt{6}\Phi}(e^{\sqrt{2/3}\Phi}-1)^2
  \label{dddp1}
\end{equation}
which shows obviously that the graviscalar potential is always negative
in the whole value of $\Phi$. Explicitly, the potential starts
with
$V_{GS}=0$ at $\Phi=0$,  decreases until it arrives at the minimum
and then again  increases  as $\Phi$ increases.
And it takes the final form of  $V_{GS}(\infty)\to 0$.
Hence the graviscalar potential on the dilatonic domain wall  is basically different
from the single Liouville-type potential of Eq.(\ref{Qpot}),
 which decreases monotonically as a scalar increases.
For the special case of  $a=0$  (non-dilatonic brane),
 we find  the RS-type potential
on the brane
\begin{equation}
V_{RS}(\Phi)=- 3k^2 e^{-\sqrt{6}\Phi}(e^{\sqrt{2/3}\Phi}-1)^2
  \label{RSp}
\end{equation}
which is also negative.
This leads to the RS vacuum solution $V_{RS}=0$ only for
the purely  graviton propagation with
$\Phi=0$. Even if we consider the dilatonic domain wall ($a\not=0$),
this does not change the negative nature of potential.
Here we have always a negative energy density
$\rho= \dot \Phi^2/2 + V_{GS}(\Phi)$ and
a positive pressure $p= \dot \Phi^2/2 - V_{GS}(\Phi)$.
 This contrasts  to the quintessence which is  based on
the positive energy density and negative pressure with
$p<-\rho/3$.
Hence even though we obtain the Liouville-type potentials,
 we cannot find any accelerating universe from the the
dilatonic domain wall including the RS-brane world scenario.
This means that the  KK scalar is not suitable for describing the
quintessence in cosmology.

\section{Dilaton as a dynamic scalar}

In this section we study the dilatonic case ($d(x)\not=0,h(x)=1)$ because we fail to
obtain an appropriate potential for the quintessence using the
graviscalar. From Eq.(\ref{act2}) one has
\begin{equation}
S_{\rm dil} = {1 \over {2 \kappa_4^2}} \int d^4 x \sqrt{-g} \left [
  R -\frac{4}{3}  (\partial d)^2 -  \frac{32k^2(\Delta +4)}{\Delta^2}\left (2e^{-ad}
 -  e^{-2ad} - 1\right ) \right
].
\label{Dact1}
\end{equation}
Introducing  $\tilde d(x)=\sqrt {3/4}
d(x)$, the above action leads to
\begin{equation}
S_{\rm dil1} = {1 \over {2 \kappa_4^2}} \int d^4 x \sqrt{-g} \left [
  R - (\partial \tilde d)^2 -  \frac{32k^2(\Delta +4)}{\Delta^2}\left
  (2e^{-a \sqrt{4/3} \tilde d} -  e^{-2a \sqrt{4/3} \tilde d} - 1\right ) \right
].
\label{Dact2}
\end{equation}
Here  we can read off its potential
\begin{equation}
V_{dil}(\tilde d)=  \frac{16k^2(\Delta
  +4)}{\Delta^2}\left (2e^{-a \sqrt{4/3} \tilde d}
 -  e^{-2a \sqrt{4/3} \tilde d} - 1\right ).
  \label{ddp1}
\end{equation}
We point out that the first term comes from the tension of the
domain wall ($\sigma_{DW}e^{-aD}$)
 and the second from the bulk potential term $(\Lambda e^{-2aD})$.
 The last term arises from the ansatz for the zero modes
 Eq.(\ref{5metric}) with $h=1$.
Further this   can be expressed as
\begin{equation}
V_{dil}(\tilde d)= - \frac{16k^2(\Delta
  +4)}{\Delta^2} \left (e^{-a \sqrt{4/3} \tilde d} - 1\right )^2.
  \label{ddp2}
\end{equation}
 It is noted  that
 $V_{dil}(\tilde d)$ decreases monotonically  until it  arrives the minimum of
 $V_{dil}(\infty) = -1$ as $\tilde d$ increases.
 But this belongs to a negative potential as is obviously shown by (\ref{ddp2}).
 Hence we find that  the dilaton cannot induce the quintessence.

\section{Discussions}
\label{sec-discussions}

We investigate the zero mode sector to the 5D dilatonic domain wall
solution for cosmological purpose.
First we study  the 4D gravity with the graviscalar on the brane.
Assuming the  FRW metric on the brane, one has a minimally coupled graviscalar with
the potential.
Although the graviscalar potential belongs to a kind of Liouville-type
potential, it remains negative in the whole value of $\Phi$.
This gives us in turn the negative energy density and positive
pressure, which contrasts to the usual quintessence  that has the
positive energy density and negative pressure.
Second we consider the 4D gravity with the  dilaton. Also  we find a
negative dilatonic potential.
Hence we cannot find the quintessence by using either the
graviscalar or the dilaton.
Furthermore the mixed potential from Eq.(\ref{act2}) takes a form
of $\sim -h^{-3}(he^{-ad}-1)^2$, which is negatively definite.
That is, even if we include both the graviscalar and dilaton, we find a
negative potential.
All of potentials which we find here
belong to the negative unstable potential. As a by-product, this implies that the
RS vacuum solution of $d=0,\Phi=0(h=1)$ may not be  a truly vacuum solution except the RS gauge.
In the limit of $k \to 0$,
we find the equation of state
 $p=\rho$ with $\omega=1$ which behaves as a stiff matter
with $\rho=p \sim a^{-6}$ for both the graviscalar and dilaton cases .

Finally we wish to comment on the result for the holographic quintessence\cite{CL}.
In that case they used the Hamilton-Jacobi equation to obtain a
single
dilatonic potential $U(\phi)=e^{b(\phi-\phi_0)}$ with $b<0$ on the brane.
Actually this corresponds to the first term of Eq.(\ref{ddp1})
that comes just from the tension of the
domain wall ($\sigma_{DW}e^{-aD}$).
Of course, a single term like this can induce an accelerating universe.
But we obtain
three terms which give totally us the dilatonic potential Eq.(\ref{ddp1}).
So it seems to exist a difference between our zero-mode approach  and holographic
approach.

\section*{Acknowledgement}
We thank to C. R. Cai and G. Kang for helpful discussions.
This work was supported in part by the Brain Korea 21
Program of  Ministry of Education, Project No. D-1123 and
 KOSEF, Project No. 2000-1-11200-001-3.

\end{document}